# Tawsol Symbols 3D – Towards an innovative Picture Exchange Communication Systems PECS

Achraf Othman
Mada Center

## 1.	Introduction

Pictograms (also called symbols) are widely used in daily life as a type of visual language, such as transportation venue signs (airport, rail station, etc.), road signs, care symbols on clothing, or direction symbols (Tijus et al., 2007). This shows how symbols can communicate information quickly and effectively. Fundamentally, symbols can be understood, regardless of the person's language or literacy skills. Therefore, people with communication difficulties may benefit from using symbols to comprehend what other people are saying, as well as to express themselves. There are many sets of pictograms available online or in the market as printed cards, some of them are free and some must be purchased. Pictogram sets can be considered in several ways including how pictorial, how guessable, how flexible, how consistent, and how visually complex. Each symbol set has strengths and weaknesses, and the choice of a symbol set should be based on the needs and abilities of the person using AAC.

Selecting pictograms for the communication environment is also important, this will include language and culture. Practical issues such as how the symbols are to be used, if software is available to produce printed materials, or which sets are available for a particular AAC device will also influence the pictogram selection. Pictograms or symbols are mostly offered as collections or sets. Most present the symbol together with the word or phrase it stands for.

Typically, the word is printed above the symbol if the focus is on communication as communication partners need to be able to see the words because they may not know what all the symbols mean. When the focus is on literacy, the reader may require seeing the symbols to help decode the written word; as emerging readers often point to words as they read, the symbol is printed above the word.

## 2.	Overview of the Tawasol Symbols

Symbol sets can be considered in several ways including how pictorial, how guessable, how flexible, how consistent, and how visually complex. Each symbol set has strengths and weaknesses, and the choice of a symbol set should be based on the needs and abilities of the person using AAC. Selecting symbols for the communication environment is also important, this will include language and culture preferences. Practical issues such as how the symbols are to be used are software available to produce printed materials, or which are available for a particular AAC device, will also influence any choice. Selecting a symbol set is predominantly based on meeting individual needs within a setting. For example, considering acceptable symbol design and communication environment. AAC users can benefit from choices of globalized, localized, and personalized symbols.

The Tawasol Symbols project's aim was to develop a freely available Arabic Symbol Dictionary suitable for use by individuals who have a wide range of communication and language difficulties and to develop a set of symbols that are culturally, linguistically, and environmentally appropriate for AAC users in Qatar and the Arab countries (Tawasol Symbols, 2020) (Figure 1). The Tawasol Symbol dictionary contains until today 1600 localized symbols.

There are many reasons for introducing a new set of localized symbols in the Arab World such as the vast differences in linguistic structures between the Arabic and English languages, which can be confusing and generate fragmented sentences, as illustrated in Figure 1.

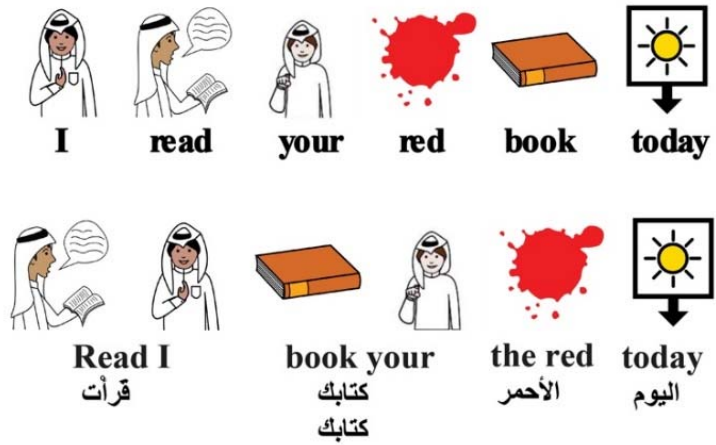

*Figure 1. Differences in structure between Arabic and English languages*

Moreover, there are always requests by teachers, therapists, and other researchers in the field for symbols not available in other languages, i.e., English, Islamic and culture-related symbols. Non-symbolic, as well as symbolic forms of communication, are culturally dependent. Hence, it is essential to customize AAC resources to meet the Arabic characteristic rubric written system and to address the presence of diglossia and the absence of culturally appropriate vocabulary.

## 3.    3D Pictograms and Augmented Reality

AR-based apps are used to enhance engagement, motivation, and learning for people with ASD. AR overlays, like 3D videos, figures, and information, can be added to anything and multiple studies have shown that these AR experiences result in increased engagement, enjoyment, motivation, and attention. The study (Yakubova et al., 2021) designed to teach object discrimination revealed a 62% increase in on-task participation and happier, more determined students. A new Google Glass-based AR and artificial-intelligence app motivates and rewards users for social and cognitive learning.

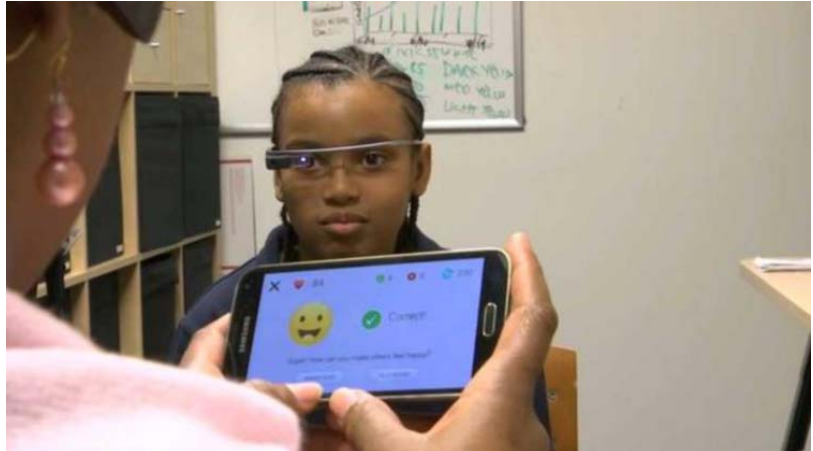

*Figure 2.Brain Power using AR to help those with autism to better connect to the world around them. Credit: Rob Michaelson (https://spellboundar.com/)*

Researchers used an AR system with foam blocks and a TV screen that acted as a mirror in order to facilitate pretend play. The foam blocks transformed into a 3D car, train, or airplane on the screen and the kids could see themselves playing with the items as toys. Results showed a significant increase in imaginative play frequency and duration with the AR scenario, and a video analysis revealed the children engaged in over 50% more pretend play scenarios per minute than without.

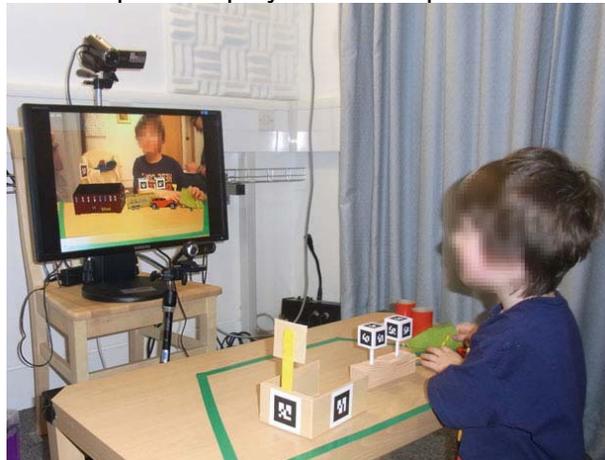

*Figure 3. The AR system designed by Zhen Bai to help children with pretend play. Photo: Graphics & Interaction Group/University of Cambridge Computer Laboratory*

In 2020, Mada Center launched a new initiative to provide researchers working on the use of Augmented Reality to improve communication skills of children on ASD. The aim of the project is designing a set of existing symbols in three dimensions. The library is useful to develop new applications using Augmented Reality technology. The 3D symbols are provided under the creative commons license. Until today, 200 3D symbols are available for download.

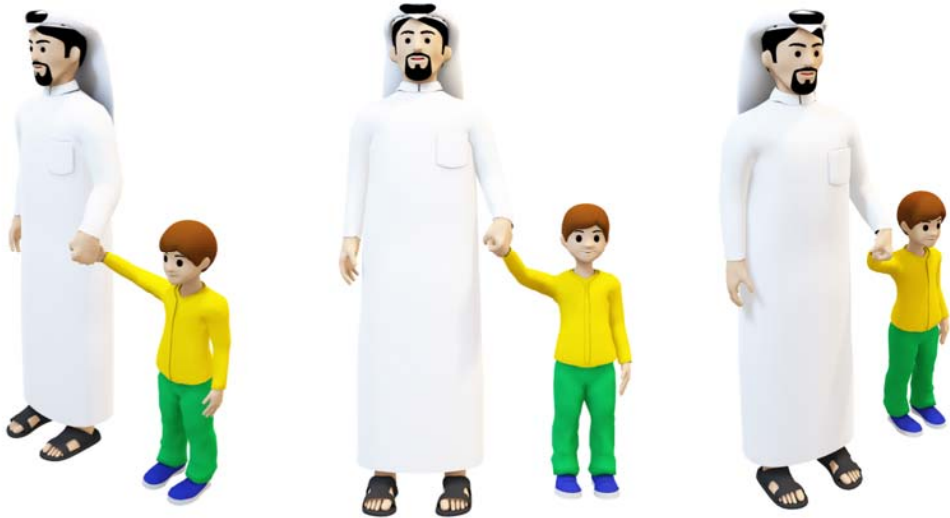

*Figure 4. An example of a 3D Tawasol Symbol from different angles*

## 4. Conclusion

AR allows for interaction with the real world which makes it easier to generalize real-life situations through digital content. The immersive, visual nature of AR capitalizes on a strength largely held by people with ASD and produces more curiosity and engagement. Introducing new technology can also be highly motivating, creating a more in-depth learning experience. In addition, AR can be easily adapted to supplement evidence-based practices, such as picture prompting and video modeling, that are currently being used by clinicians.

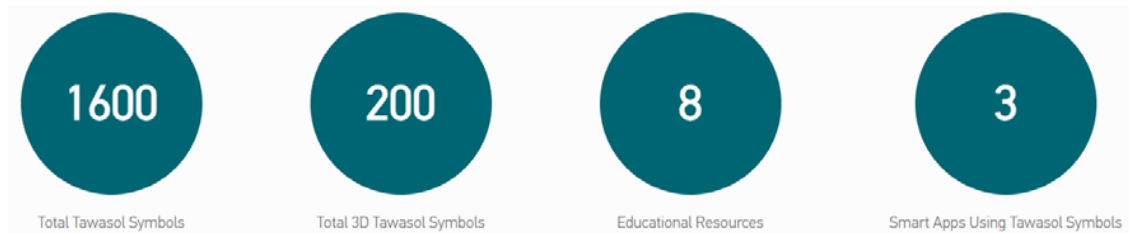

*Figure 5. Stats of the Tawasol Symbol Project*

The Tawasol Symbols project developed and localized 1600 symbols and 200 3D symbols, however, there are still areas of improvement such and the knowledge around users' priorities and core and fringe vocabulary. More research is encouraged to develop Arabic symbols and to investigate suitable interactive technology devices that use AAC in a bilingual setting. This project opens the door for research opportunities to cultivate a more effective localized and personalized communication system.